\documentclass{article}
\usepackage[round, sort&compress, numbers]{natbib}
\usepackage{setspace}
\usepackage[margin=1in]{geometry}
\usepackage{graphicx} 
\usepackage{subcaption}
\usepackage{amsmath}
\usepackage{url}
\usepackage{datetime}
\usepackage{hyperref}
\usepackage{multirow}
\usepackage{array}
\usepackage{float}
\usepackage{xcolor,colortbl}
\usepackage{siunitx}
\usepackage{authblk}

\newdateformat{mydateformat}{\monthname[\THEMONTH] \THEDAY, \THEYEAR}

\title{Audit of takedown delays across social media reveals failure to reduce exposure to illegal content}

\author[1,2]{Bao Tran Truong}
\author[1]{Sangyeon Kim\thanks{Corresponding author. Email: clearingkim.research@gmail.com}}
\author[3]{Gianluca Nogara}
\author[3]{Enrico Verdolotti}
\author[3,4]{Erfan Samieyan Sahneh}
\author[5]{Florian Saurwein}
\author[5]{Natascha Just}
\author[6]{Luca Luceri}
\author[3]{Silvia Giordano}
\author[1]{Filippo Menczer}

\affil[1]{Observatory on Social Media, Indiana University, Bloomington, USA}
\affil[2]{Center Synergy of Systems, Dresden University of Technology, Germany}
\affil[3]{Dept.~of Innovative Technologies, 
University of Applied Science and Arts, Switzerland}
\affil[4]{Dept.~of Computer Science and Engineering, University of Bologna, Italy}
\affil[5]{Media \& Internet Governance Division, IKMZ, University of Zurich, Switzerland}
\affil[6]{Information Sciences Institute, University of Southern California, USA}

\date{}
\newcommand{\appenx}{Appendix}
\newcommand{\simsom}{\emph{SimSoM}}

\begin{document}

\maketitle

\begin{abstract} 
Illegal content on social media poses significant societal harm and necessitates timely removal. 
However, the impact of the speed of content removal on prevalence, reach, and exposure to illegal content remains underexplored.  
This study examines the relationship with a systematic audit of takedown delays using data from the EU Digital Services Act Transparency Database, covering five major platforms over a one-year period. 
We find substantial variation in takedown delay, with some content remaining online for weeks or even months. 
To evaluate how these delays affect the prevalence and reach of illegal content and exposure to it, we develop an agent-based model and calibrate it to empirical data. 
We simulate illegal content diffusion, revealing that rapid takedown (within hours) significantly reduces prevalence, reach, and exposure to illegal content, while the longer delays measured by the audit fail to reduce its spread. 
Though the link between delay and spread is intuitive, our simulations quantify exactly how takedown speed shapes exposure to illegal content. 
Building on these results, we point to the benefits of faster content removal to effectively curb the spread of illegal content, while also considering the limitations of strict enforcement policies. 
\end{abstract}

\section*{Introduction}

Illegal and harmful content --- including copyright violations, child sexual abuse material (CSAM), and incitement to violence or terrorism --- poses serious harm and demands effective moderation efforts \cite{yar2018failure, jain2020illegal}. 
Platforms have developed a broad range of moderation tools and strategies that vary in severity.
For some problematic content, such as misinformation and hate speech, interventions have been designed to promote critical user engagement, such as sensitive content labels \cite{morrow2022emerging, martel2023misinformation}, accuracy reminders \cite{bhuiyan2018feedreflect, pennycook2020fighting, pennycook2021shifting, pennycook2022accuracy}, and friction mechanisms in content sharing \cite{jahanbakhsh2021exploring, jahn2023friction, tomalin2023rethinking}. These ``soft'' interventions alter the environment to nudge user behaviors and are thus contingent on user motivations. Other techniques reduce the visibility of problematic content by limiting its searchability, recommendation, and engagement \cite{gillespie2022not, macdonald2024moderating, saurwein2025beyond}.
Stricter interventions include content removal (takedown), account suspensions, and permanent bans \cite{chandrasekharan2017you, ali2021understanding, jhaver2021evaluating}. 
These approaches enable swift disruption of content virality and minimize exposure, but also raise concerns about censorship \cite{jiang2023trade}. 

Given the trade-offs of different moderation approaches, previous modeling work has shown that combining interventions is an effective approach to leverage the strengths of each while mitigating their limitations \cite{bak2022combining}. Nevertheless, for clear-cut cases of illegal content, removal or deactivation in concerned jurisdictions remains a legal demand for social media platforms. 
The importance of timely content removal has also been recognized in regulatory frameworks. Laws such as Germany's Network Enforcement Act (NetzDG), Austria's Communication Platform Act (KoPl-G), and France's Loi Avia required platforms to remove ``obviously illegal'' content within 24 hours and other illegal content within seven days. These provisions reflected a clear regulatory emphasis on speed and have drawn criticism both for not being strict enough and for being too strict --- potentially infringing on free expression  \cite{Kozyreva2023dilemmas, schulz2019probleme, heldt_2020, vie_publique_2020}. 
One of the critical issues is the trade-off between moderation speed and accuracy: removing content very quickly can reduce its spread but may also lower precision, causing false-positive errors and legitimate concerns about over-moderation. 
The Digital Services Act (DSA) replaced precise deadlines with broader requirements: platforms must act ``in a timely manner'' (Art.~16(6)) and remove or disable access to illegal content ``expeditiously'' (Art.~6(1)). While such flexibility avoids rigid constraints, it also marks a shift away from legally enforceable time limits. 

As legal requirements have shifted, the question of how takedown speed affects the prevalence, reach, and exposure of illegal content becomes highly relevant. However, thus far, there is no empirical data on the effects of takedown speed, making it difficult for policymakers to design and evaluate appropriate policies. 
This paper provides a comprehensive evaluation of the effects of moderation speed  by proposing a model that measures how prevalence, exposure, and reach of illegal content change in response to takedown delay. In addition, we audit available data on takedown delay of selected platforms and discuss the consequences and limitations of regulatory takedown deadlines.

Available research on the speed of content removal shows that delay in response reduces the moderation effectiveness on prevalence and exposure \cite{schneider2023effectiveness, goldstein2023understanding, edelson2025measurement}. However, the existing studies largely focus on mis- and disinformation, such as misleading claims or conspiracy theories, which may not generalize to illegal content. Moreover, much of this work centers on emergencies, such as COVID-19 anti-vaccination narratives \cite{broniatowski2022evaluating} or the January 6 Capitol Riot \cite{goldstein2023understanding}, during which platforms adopted exceptional measures. To inform legal frameworks, it is necessary to examine the standard operating procedures of platforms during routine, non-emergency periods. Studying the impact of content removal policies at scale is challenging, both because it would be ethically impermissible to experiment with harmful or illegal content, and because provision of access  to illegal content  is critical and involves privacy concerns. Currently, no studies systematically examine illegal content removal practices across platforms or quantify the effects of takedown delays. 

Based on a year of data from the \textit{DSA Transparency Database} (DSA-TDB) \cite{dsadatabase}, we show that illegal content removal speed varies widely across platforms, ranging from several days to weeks, and in some cases, up to a year. By calibrating an agent-based model with empirically estimated takedown delays, we quantify the relationship between moderation speed and the reduction of illegal content spread. 
While longer delays predictably result in greater prevalence, reach and exposure of illegal content, our simulations reveal the precise, non-linear effect of takedown speed. The results indicate a steep increase in prevalence and exposure as takedown delay increases. Prompt removal of illegal content markedly reduces spread compared to systems with no or slow moderation. The impact of content removal becomes negligible as the delay extends beyond a couple of weeks --- a timeframe that, according to our data, is typical for the takedown of illegal content on most platforms.

Our findings indicate that faster content removal can significantly reduce the spread of illegal material. However, current takedown delays often allow such content to circulate widely before action is taken. This suggests that regulatory frameworks like the DSA may benefit from stronger enforcement provisions aimed at promoting more timely and effective moderation. However, in the \nameref{sec:discussion}, we point to limitations of takedown deadlines and argue that any legal framework should balance the goal of timely removal with the need to ensure the quality of moderation and account for platform-level operational constraints. 

\begin{figure}
\centering
\includegraphics[width=\linewidth]{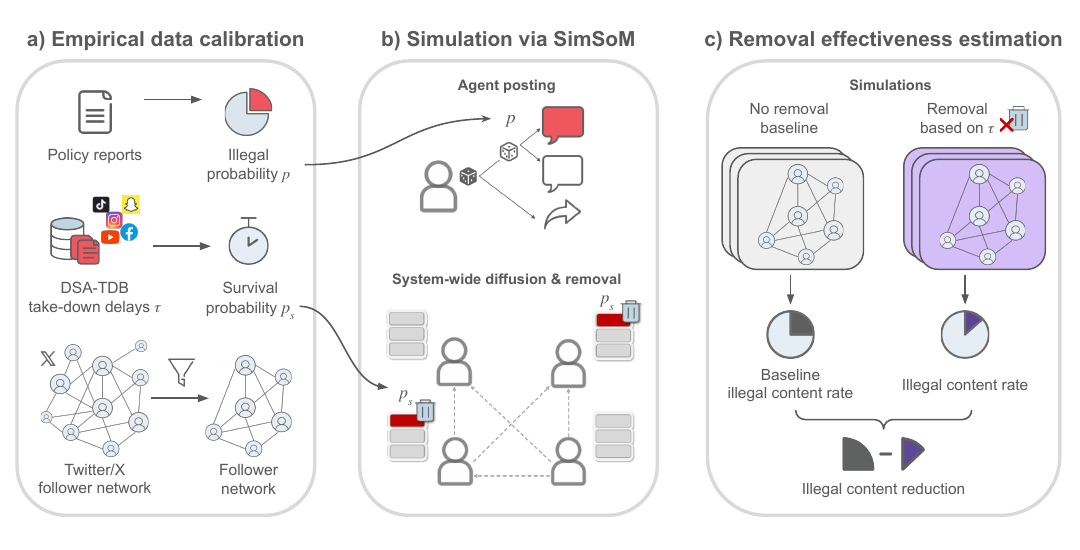}
\caption{\textbf{Pipeline for estimating the effectiveness of illegal content removal.}
(a)~Empirical data calibration: We estimate a plausible range for the probability of illegal content $p$ from policy reports and infer platform-specific takedown delays from Statements of Reasons (SoRs) in the DSA Transparency Database (DSA-TDB) for five major platforms: Facebook, Instagram, YouTube, TikTok and Snapchat. These delays are summarized by the expected content takedown delay $\tau$ and used to parameterize content removal. Simulations are run on an empirical follower network derived from Twitter data.
(b)~Information diffusion and removal: An agent-based model simulates the spread of content over the network. At each timestep, illegal content survives with constant probability $p_s$, calibrated from the empirical takedown delay $\tau$; once removed, content is deleted from all user feeds.
(c)~Moderation effectiveness estimation: Simulations are performed across a broad range of takedown delays to quantify the impact of removal speed. The effect of content removal is measured as the relative reduction in the prevalence and exposure of illegal content, compared with a baseline without removal.
}
\label{fig:schema}
\end{figure}

\section*{Methods}
\label{sec:methods}

We begin by analyzing the DSA-TDB to understand the illegal content removal practice on different platforms.
We then introduce a three-step pipeline to evaluate the effectiveness of moderation under different time delays using an agent-based model, as outlined in the following sections and illustrated in Fig.~\ref{fig:schema}. 
We first describe the empirical estimation of the parameters used to calibrate the simulations. These include (i)~the probability that newly created content is illegal, estimated from external policy reports; and (ii)~the expected takedown delays of illegal content, derived from the DSA-TDB. 
Next, we describe the extensions made to  \simsom, an agent-based model of social media, to capture information diffusion and content removal mechanisms. Finally, we detail the experimental setup used to run multiple simulation scenarios.

\subsection*{Estimating takedown delay from the DSA Transparency Database}
\label{sec:dsa_calibration}

To study cross-platform takedown speed for illegal content, we rely on the \textit{DSA Transparency Database (DSA-TDB)} \cite{dsadatabase}, which provides case-level Statements of Reasons (SoRs) with timestamps for content publication and removal, allowing us to measure how long content remained online. 
Other available takedown data come from platform \textit{Transparency Reports}, for example under Art.~15 of the DSA. They typically report aggregated notice-to-action times, which capture responsiveness after notification but not the publication-to-removal duration required to measure how long content was online and potentially reachable by users, which is central to our analysis (see \appenx, \nameref{sec:transparency_reports}). 

We collect SoRs from the DSA-TDB between January 1 and December 31, 2024. We capture relevant records using the strictest filter, where the platforms explicitly state that the moderation is due to illegal content.\footnote{The DSA-TDB contains reporting fields to indicate the decision ground for content moderation. Content must be classified either as illegal (DECISION GROUND ILLEGAL CONTENT), or as incompatible with terms and conditions (DECISION GROUND  INCOMPATIBLE CONTENT). For our analysis, we select SoRs that classified content as illegal. However, since platforms do not have do conduct legal assessments of all moderated content, more than 99\% of content is classified as incompatible with terms and conditions, and the share of illegal content may be underreported \cite{trujillo2025dsa,groesch2025big}. In addition, platforms can indicate in an optional boolean field if content, which was classified as incompatible with terms and conditions, is also considered illegal (INCOMPATIBLE CONTENT ILLEGAL). Since platforms do not use this reporting category consistently, we did not include it in our dataset for calibration.}
We consider five major social media platforms with a sufficient number of relevant SoRs: Facebook, Instagram, YouTube, TikTok, and Snapchat.
We exclude LinkedIn and Pinterest due to a lack of SoRs related to illegal content and Twitter/X due to suspicious patterns revealed in its DSA data, as also highlighted in prior work \cite{trujillo2025dsa, groesch2025big}. 
This results in 8,613 SoRs submitted by Facebook, 3,005 by Instagram, 408,400 by YouTube, 2,142 by Snapchat, and 90,409 by TikTok.\footnote{We excluded 2,949 TikTok observations with invalid input for the content creation date (2000-01-01).} 
The takedown delay is defined as the difference between the moderation date (\textit{application\_date} field) and the content posting date (\textit{content\_date} field). 

Beyond highlighting variation in content removal practices across platforms, this data source provides a way to empirically estimate the \emph{expected takedown delay} of illegal content. Throughout the paper, we denote this quantity by $\tau$. Empirically, $\tau$ summarizes observed moderation behavior in the DSA-TDB; in the simulation, $\tau$ serves as a parameter governing the stochastic removal process.

\subsection*{Estimating the probability of illegal content}

We define the \emph{illegal content probability} $p$ as the mean prevalence of illegal content in the system. Empirically estimating $p$ is challenging, as platforms rarely provide access to original content or systematic assessments of legality. As a result, there is limited direct evidence on the prevalence of illegal content in large-scale online systems (for an exception, see Wagner et al.\cite{wagner2024mapping}). To estimate a plausible range for $p$, let us analyze the Future of Free Speech policy report~\cite{futurefreespeech2024}. The report includes comment data collected from Facebook and YouTube across Germany, France, and Sweden during two weeks in June--July 2023. 
It also includes the fraction of comments that disappear ($\frac{\text{disappear}}{\text{total}}$) and the proportion of disappeared comments that are legal ($\frac{\text{legal}}{\text{disappear}}$). From this, we estimate the fraction of comments that are illegal: $\frac{\text{illegal}}{\text{total}} = \frac{\text{disappear}}{\text{total}} \times \frac{\text{illegal}}{\text{disappear}} = \frac{\text{disappear}}{\text{total}} \times (1 - \frac{\text{legal}}{\text{disappear}})$. 
This approach treats all disappeared comments as potential removals, regardless of whether deletion was performed by the platform, moderators, or the original author. As a result, it may overestimate the true prevalence of illegal content, since some removed content is later classified as legal. Nevertheless, it provides a conservative upper bound. 
Table~\ref{table:futurefreespeech_illegalprob} shows that this approach yields a wide range between 0.00002--0.009.
We run simulations across this range of $p$ values, excluding $p=0.00002$ that would require infeasibly large networks to produce stable estimates. We find that system-level metrics are largely independent from the value of $p$ (see \appenx, \nameref{sec:robustness}); we use $p=0.01$ in the simulations.  

While $p$ captures the mean prevalence of illegal content, it does not determine how such content is distributed across users. In practice, illegal content may be generated uniformly across the population or be highly concentrated among a small subset of accounts. We examine several alternative distributions of individual-level illegal posting probabilities that all share the same population mean $p$: (i)~a near-homogeneous scenario in which all users have similar propensities to post illegal content; (ii)~a heterogeneous scenario in which some users have substantially higher probabilities of posting illegal content than others; and (iii)~a two-group scenario in which the population consists of a small fraction of high-risk users who repeatedly post illegal content and a majority of low-risk users who only occasionally violate the law (see Appendix, \nameref{sec:illegal_prob}). 
Results are robust across all three assumptions about how illegal content probability is distributed across users (see \appenx, \nameref{sec:robustness}). 
We decided to model illegal content as being generated by two groups, each characterized by a beta distribution (see \appenx, \nameref{sec:illegal_prob}). 

\begin{table}
\centering
\caption{Data from the Future of Free Speech policy report and derived ratios of illegal content.}
\begin{tabular}{lccl}
\hline
 & Disappeared/Total & Legal/Disappeared & Illegal Ratio \\
\hline
Germany (FB) &  0.006 & 0.997 & 0.00002 \\
Germany (YT) &  0.115 & 0.989 & 0.001 \\
France (FB) &  0.012 & 0.921 & 0.0009 \\
France (YT) &  0.072 & 0.875 & 0.009 \\
Sweden (FB) &  0.005 & 0.946 & 0.0003 \\
Sweden (YT) &  0.041 & 0.946 & 0.002 \\
\hline
\end{tabular}
\label{table:futurefreespeech_illegalprob}
\end{table}

\subsection*{Modeling social media diffusion}

We use \simsom~\cite{truong2023quantifying}, an agent-based model that mimics information diffusion on social networks such as Twitter/X, Mastodon, Bluesky, Threads, or Instagram. The platform is represented as a directed network, where nodes correspond to user accounts and edges denote follower relationships, which need not be reciprocal. 
As in real-world platforms, information circulates through messages that appear on user news feeds. At each time step, a user can post new original messages, reshare content from their feed, or take no action. Messages represent information that could take the form of text, links, hashtags, images, or other media.

To select a message for resharing, an agent samples from the news feed, which contains distinct messages shared by their friends. Once shared by an agent, a message appears in the news feeds of their followers, allowing for propagation across the network (see \nameref{sec:discussion}). Thus, the exposure of the message depends on its author's follower count --- agents with more followers will achieve greater exposure. 
The model captures mechanisms of real-world ranking algorithms and sharing behavior, according to which agents prefer to reshare messages that are appealing, recent, and popular \cite{twitteralgo, truong2023quantifying}. 

Building on the base model, we incorporate additional features to examine the diffusion of illegal content. Specifically, each message includes a binary attribute indicating whether the content is illegal. 
Different types of illegal content may be more or less appealing. For example, CSAM is criminally and morally condemned, therefore resharing is expected to be rare. 
More common forms of illegal content, such as copyrighted media, tend to be highly engaging. 
Lacking data on the correlation between legality and appeal, we make the simplifying assumption that the two are independent. 

We also extend \simsom{} to model the heterogeneous activity of social media users, where most agents have low activity levels, and only a few are hyperactive. We model this by assigning an activity level $a$ to each user at the beginning of the simulation. We draw this activity from a power law $P(a) \sim a^{\gamma}$. 
We estimate the exponent $\gamma = 2.85$ by fitting empirical measurements of average daily Twitter activity, based on 81{,}633{,}118 posts by 37{,}882{,}999 accounts collected over nine days in January 2023. 
If $a<1$, the user (re)shares a single message with probability $a$ at each time step; else the user (re)shares $\lfloor a \rfloor$ messages. 

\subsection*{Modeling illegal content diffusion and removal}
\label{sec:modeling-removal}

To understand the dynamics of illegal content removal, let us consider the lifespan of such content on a platform. While some illegal content is immediately detected and removed using automated detection and decision systems, other is missed due to detection limitations. In such cases, content moderation is delayed until the platform either detects the content itself or receives an external notification. Additional delays may occur when platforms assess the content against their community standards and national laws. If a violation is confirmed, either the content is removed globally or access is restricted in certain jurisdictions. 

While real-world moderation involves multiple detection and decision stages, our objective is to capture the aggregate lifetime during which illegal content can diffuse through the network. To this end, we model removal using a stochastic survival process. 
We make the simplifying assumption that at any time step, illegal content survives with probability $p_s$ and is removed with probability $1 - p_s$. 
Under this assumption, the probability that content survives for at least $t$ time steps is $P(t) \approx p_s^t$, since the content must survive $t$ consecutive removal opportunities. 

Empirically, takedown delays are well described by an exponential survival curve,
$P(t) \approx e^{-t/\tau}$, 
where $\tau$ is the \emph{expected takedown delay} or mean lifetime obtained by fitting the DSA-TDB data (see \nameref{sec:dsadelay}). To ensure that the simulation mirrors the same survival curve and $\tau$ values as observed in the empirical data, we parameterize the discrete-time survival process to match this exponential form at integer time steps. Since $P(t) \approx p_s^t$ in the simulation, we set $p_s = e^{-1/\tau}$.
Thus, at each time step we remove content with probability $1-e^{-1/\tau}$. When a message is removed, it is deleted from all user feeds. 

\subsection*{Simulation framework and metrics}

Having specified the diffusion and removal dynamics of illegal content, we now turn to the simulation framework used to evaluate the effects of removal delays --- the network structure, outcome metrics, convergence criterion, and parameter settings used in the simulations. 

Simulations are conducted on a follower network derived from empirical Twitter data. The network connects users obtained from a 10\% random sample of public tweets~\cite{Nikolov2020dataset}. The dataset comprises users who posted links to news articles, with likely automated accounts excluded. The original follower network constructed from this dataset has $58{,}296$ nodes and $10{,}632{,}023$ edges. 
We reduce the size of this network to speed up the simulations reported in \nameref{sec:main_results}. We employ $k$-core decomposition to identify $N = 10{,}006$ nodes that constitute the $k = 94$ core. Subsequently, a random selection of edges is removed to match the density of the original network. This procedure produces $E = 1{,}809{,}798$ edges, with each node having an average of roughly 180 friends or followers. The main results are robust with respect to the structure of the network (see \appenx, \nameref{sec:robustness}). 

We measure the effects of time delay in content takedown by the reduction of illegal content throughout the network, which can be quantified by multiple metrics. 
\emph{Illegal content prevalence} is defined as the average ratio of illegal content in circulation at time $t$: 
\(
I_t = \frac{1}{N}\sum_{i=1}^{N} f_{i,t}
\)
where $f_{i,t}$ is the fraction of illegal content in user $i$'s feed at time $t$ and $N$ is the number of agents.
The more illegal content is present in the system, the more harm it can potentially cause. 
However, a piece of illegal content may be removed from a user's feed before the user sees it. Therefore we also measure potential impact more directly using two exposure metrics. 
\textit{Impressions} measures the number of times illegal content is seen by users. Note that a single user can be exposed to many posts with illegal content, or even the same post multiple times. 
\textit{Reach} is a complementary measure of exposure, capturing the number of unique users who see any amount of illegal content. 
For each of these metrics, we calculate the reduction percentage relative to a baseline measure without any content removal. For example, the reduction in illegal content prevalence is calculated by $(\langle \bar{I}_{\text{baseline}} \rangle - \bar{I}_{\text{removal}}) / \langle \bar{I}_{\text{baseline}} \rangle$, where $\bar{I}$ is the illegal content prevalence at the end of a simulation and $\langle \bar{I}_{\text{baseline}} \rangle$ is the baseline average. 

The simulation ends once the system reaches a \emph{steady state}, signaled by a convergence in the illegal content prevalence. To this end, we calculate an exponential moving average at time $t$ as $\bar{I}_t = \rho \bar{I}_{t-1} + (1-\rho) I_{t}$, where $\rho$ is a parameter regulating the importance of older values.   
We stop the simulation when the difference between two consecutive values of this moving average is smaller than a threshold, $|\bar{I}_t - \bar{I}_{t-1}| / \bar{I}_{t-1} < \epsilon$. 
The parameters $\rho =0.9$ and $\epsilon = 0.0001$ used in the reported simulations were tested to ensure that the system's illegal content prevalence stabilizes at the steady state. 

For each set of parameters, we run 70 simulations without any content removal; these are averaged to obtain baseline metrics. 
We then run 70 simulations with content removal; illegal content prevalence and exposure metrics (reach and impressions, defined in \nameref{sec:main_results}) are averaged across these runs. 
All simulations start from random conditions. 

\section*{Results}

\subsection*{Long takedown delays across platforms}
\label{sec:dsadelay}

\begin{figure}[t]
\centering
\includegraphics[width=\linewidth]{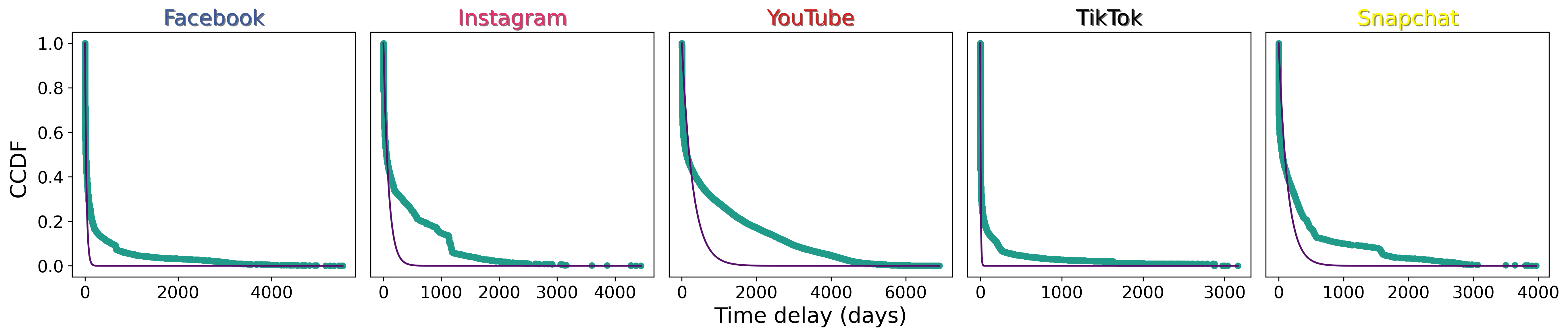}
\caption{
Complementary cumulative distribution functions (CCDFs) of time delay in illegal content takedown on five platforms. Each plot reports empirical data from the DSA-TDB (green dots) along with an exponential fit (purple line). 
}
\label{fig:timedelay_curve}
\end{figure}

Our analysis of the DSA-TDB for five major social media platforms reveals substantial variation in illegal content takedown delays, ranging from several days to weeks and in some cases up to several years. 
Figure~\ref{fig:timedelay_curve} presents the empirical distributions of takedown delays derived from DSA-TDB Statements of Reasons for the five platforms. 
By fitting exponential survival models of the form $P(t) \sim e^{-t/\tau}$ to these empirical curves, we estimate expected takedown delays $\tau \approx 31$ days for Facebook, 87 days for Instagram, 286 days for YouTube, 6 days for TikTok, and 136 days for Snapchat. 
The exponential models provide a reasonable approximation for short delay periods but diverge from the empirical curves at longer times, where heavier tails are evident. 
Consequently, these estimates of $\tau$ should be interpreted as optimistic summaries, as the exponential specification downweighs rare but prolonged takedown delays observed in the data. To account for both cross-platform variation and uncertainty in the tail behavior of moderation delays, our simulations (Section~\nameref{sec:main_results}) explores a broad range of $\tau$ values that both encompasses and extends beyond the empirically observed estimates, up to a maximum of $\tau = 739$ days.

\begin{table}
\centering
\caption{Expected content takedown delay $\tau$ of illegal content by type and platform.}
\begin{tabular}{lrrrr}
\hline
 & \multicolumn{2}{c}{YouTube} & \multicolumn{2}{c}{TikTok} \\
\hline
Illegal Content Type & Observations & Delay & Observations & Delay \\
\hline
Copyright violation & 375,531 & 276  & --     & -- \\
Privacy violation & 17,737  & 2,362 & 14,174 & 8 \\
Other illegal & 9,076   & 58   & 15,375 & 7 \\
Counterfeit & 2,500   & 78 & --     & -- \\
Circumvention & 1,174   & 769  & --     & -- \\
Hate speech & 763 & 12 & 19,036 & 10 \\
Defamation & 718 & 36 & 1,225  & 5 \\
Child endangerment & 647 & 88   & 11,881 & 1  \\
Court order & 254 & 174 & --     & -- \\
Non-consensual intimate imgs & --      & --   & 12,271 & 6 \\
Fraud & -- & -- & 4,922 & 54 \\
Harassment & -- & -- & 4,853  & 4 \\
Terrorism & -- & -- & 4,158  & 8 \\
National security & -- & -- & 1,265 & 47 \\
CSAM & -- & -- & 1,249 & 5  \\
\hline
\end{tabular}
\label{table:illegal_content_delay_platform}
\end{table}

Variation in takedown delays within platforms may further reflect differences across types of illegal content, as some violations can be assessed quickly while others require more extensive investigation. 
Table~\ref{table:illegal_content_delay_platform} reports estimated $\tau$ values by illegal content type for YouTube and TikTok. 
Corresponding empirical survival curves for each content type are shown in \appenx, \nameref{sec:by_type}. 
The remaining three platforms are excluded from this analysis because their SoRs do not provide sufficiently informative classifications of illegal content types. 
Both YouTube and TikTok exhibit substantial heterogeneity across content categories: on YouTube, estimated delays range from 12 to 2,362 days, whereas on TikTok they range from one to 54 days. 
While this variation underscores that illegal content moderation is far from uniform, we next simulate a range of takedown scenarios designed to encompass most of the empirically observed values of $\tau$.

\subsection*{Takedown fails to curb illegal content prevalence and exposure}
\label{sec:main_results}

Based on the above empirical analysis, we simulate various takedown scenarios by exploring a broad range of expected content takedown delay values, 2 hours $\leq  \tau \leq 739$ days.
We measure the effects of time delay in content takedown by the reduction of illegal content throughout the network, as quantified by the prevalence and exposure metrics defined in \nameref{sec:methods}. 

\begin{figure}
\centering
\includegraphics[width=0.7\linewidth]{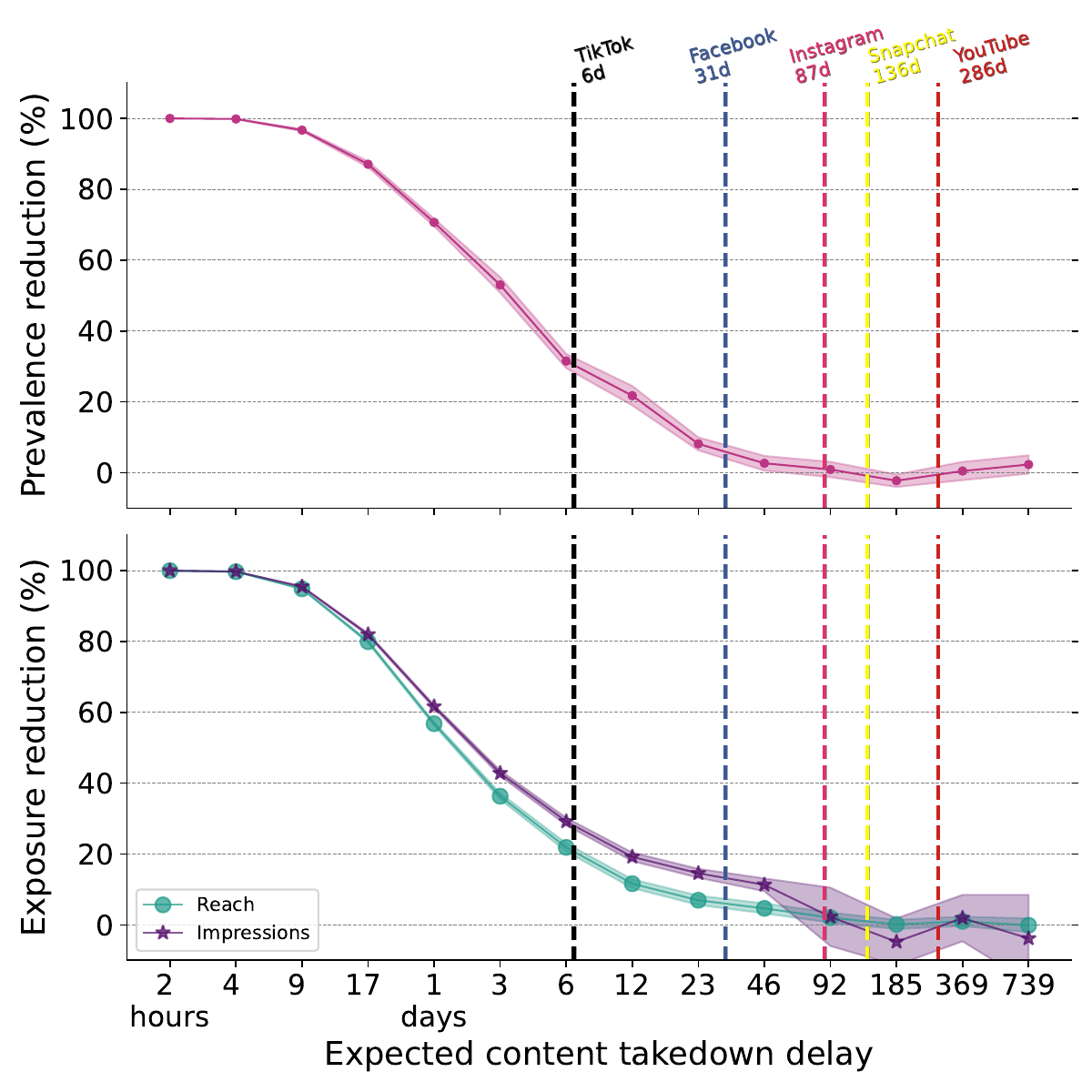}
\caption{Impact of takedown delay on illegal content. The reduction in the prevalence (pink), reach (green), and impressions (purple) of illegal content is plotted as a function of the expected takedown delay $\tau$. 
Shading represents the 95\% confidence intervals estimated using non-parametric bootstrapping.  
The vertical lines indicate the expected takedown delays obtained by fitting the DSA-TDB data for different platforms. 
}
\label{fig:res_frac}
\end{figure} 

Fig.~\ref{fig:res_frac} plots the reduction in illegal content prevalence, impressions, and reach as a function of the expected takedown delay.
Content removal has maximal impact under rapid takedowns. For expected delay below nine hours, removal results in approximately 95--100\% reduction of illegal content across all measures. 
Moderation becomes less effective as the expected time delay increases, allowing illegal content to spread before being removed. 
Between approximately 9 hours--46 days, the reduction in prevalence of illegal content drops from 95\% to negligible levels. 
The empirical values measured from DSA-TDB data for TikTok and Facebook are within this interval. 
Those for Instagram, Snapchat, and YouTube correspond to longer delays, where removal has no significant effect on illegal content prevalence. 
Although the overall trend is similar across the two exposure metrics, the reduction in impressions and reach drops even faster for delays of up to a few days. 
For example, when illegal content is removed within a day on average, reach and impressions are only reduced by about 60\%.
In summary, these findings show that delays beyond a day significantly diminish the effect of content removal, and that all platforms have longer delays, even in our optimistic estimates.  

\section*{Discussion}
\label{sec:discussion}

This paper reveals how delays in illegal content takedown affect prevalece and exposure of illegal content. Using empirical takedown data for five social media platforms from the DSA-TDB, we find large variations in average takedown delays. While TikTok typically responds within a week, illegal content often remains online for over a month on Facebook, Instagram, Snapchat, and YouTube.
Our agent-based model calibrated on this data shows that such long delays can substantially diminish the impact of content removal. 
Although this relationship may appear intuitive, our analysis reveals non-linear dynamics between takedown speed and moderation effectiveness, including critical delay thresholds.
Specifically, we show that removing illegal content within one day significantly reduces its prevalence and exposure, whereas prolonged delays allow illegal content to permeate user feeds --- beyond 46 days, takedown becomes futile in curbing the spread of illegal content. 

These findings have practical implications for legislation and social media moderation. 
For legislation, the importance of timely moderation appears to support the argument for implementing tight illegal content removal deadlines, as imposed under NetzDG in Germany and KoPl-G in Austria prior to the DSA. 
However, tight takedown deadlines may have detrimental effects on moderation quality and freedom of speech. 
Takedown deadlines reduce the time necessary to properly assess the (un)lawfulness of content, which often requires careful verification, including report reviews and appeal resolutions.  
Furthermore, definitions of what content is illegal differ across countries, demanding careful assessment of national context to avoid over-removal. 
On the other hand, legislation that defines tight takedown deadlines may impose fines for platforms that systematically fail to meet them. 
The threat of such fines may incentivize platforms to over-remove content to avoid sanctions, thereby stifling freedom of speech online \cite{CCIAEurope2017}. 
In France, the Loi Avia was struck down by the Constitutional Council \cite{conseil_constitutionnel_2020} precisely because of concerns about restrictions on freedom of expression \cite{vie_publique_2020,heldt_2020}. The Constitutional Council argued that the determination of the unlawfulness of the content is often not based on its obvious nature but is a matter for assessment by competent authorities. The stringent takedown deadline granted to the platforms for enforcement did not allow operators to wait for a decision by a judge, thus violating freedom of expression. Any regulatory framework must ensure effective moderation of illegal content while remaining practical for platforms, and mindful of potential adverse effects. 

These considerations temper the implications of our results for proactive moderation policies. 
Even though reducing takedown delays could significantly limit the spread of illegal content, platforms should remove immediately only obvious cases. 
For cases requiring further assessment (e.g., borderline content, national legislation requirements), platforms should take time for thorough review to avoid hasty decisions and over-removal. Until the assessment is completed, platforms may implement alternative interventions, such as downranking or reduced visibility, to reduce the circulation of potentially illegal content without resorting to premature removal. Another approach to improving content moderation is to strengthen collaboration with trusted flaggers --- individuals or organizations that identify and report content violating community guidelines or legal standards. With human flaggers being a limited resource \cite{fbdsa24oct}, efforts should be directed towards prioritizing the most time-sensitive cases.

Another issue arising from strict regulatory takedown deadlines is that they can only address the timespan between the notification of illegal content and its removal, not the period during which illegal content circulates in the network undetected. Therefore, the impact of tight deadlines on the spread of illegal content is inherently limited. In fact, most platforms already report fast reactions after notifications: a median time below 24 hours according to 2024 DSA Transparency Reports and removal of 80\% of illegal content within 24 hours according to 2022 national Transparency Reports for Germany and Austria.  These figures suggest that a large portion of the long delays measured from the DSA-TDB data occurs before illegal content is detected or flagged. 

To further increase the utility of the DSA Transparency Database for research and policy on content moderation, platforms should be required to report the timestamps not only for publication and removal of content, but also for detection/notification. This would make it possible to identify how long content has circulated in the network undetected, and to evaluate if platforms do react timely to violations and notifications. Moreover, SoRs for content and account restrictions should include additional indicators of reach and scale. Simple engagement metrics (number of followers, views, reshares) would enhance interpretability and support monitoring the spread of illegal content. Finally, individual SoRs should be linked to accounts in a privacy-preserving way to enable identification of the volume of repeat offenders \cite{groesch2025big}.

While our findings are robust, several limitations warrant consideration. 
The lack of empirical data on illegal content results in several assumptions about the content production process in our model. 
First, the model focuses on the spread of content already posted, without accounting for the user incentives or legal risks influencing decisions to post illegal material. 
Second, we assume static probability distributions of sharing such content. Future research could extend our model by accounting for the fact that users often alter their posting patterns in response to moderation decisions, potentially declining under stricter or faster moderation \cite{srinivasan2019content, jhaver2019does}.  
Third, as mentioned in \nameref{sec:methods}, we assume uniform appeal and spreading potential across all types of content. If illegal content is generally more engaging, the model may underestimate the impact of delayed takedown, and vice versa. 
Fourth, we triangulate data from multiple sources, which may introduce inconsistencies. For instance, user activity is calibrated using global Twitter data (in particular tweets linking to news), whereas the DSA-TDB focuses on European users. This mismatch could lead to unknown biases in our calibration.
Lastly, our extension of \simsom neglects out-of-network content that could be recommended in user feeds. Further experiments could explore the effects of out-of-network recommendations on illegal content prevalence and exposure. 

Our model of the moderation process also has limitations. The DSA-TDB data only captures the downstream outcomes of content moderation, neglecting the detection process. 
The model thus treats all illegal content as equally detectable, which may overlook important variation in the detection of different content types. 
We also make the simplifying assumption that when illegal content is removed, it is deleted from all user feeds. In practice, however, some content may be illegal only in certain jurisdictions and access may be selectively disabled for users in these countries. 
We do not consider such a differentiation. 

The database itself has limitations \cite{trujillo2025dsa,groesch2025big}. It relies heavily on self-reporting by platforms, leading to inconsistent coverage and reporting quality both across and within platforms. This highlights a broader issue with self-reporting systems. Moderation decisions are shaped by platform-specific moderation guidelines and practices. Classification of content as illegal does not follow coherent and reliable procedures across all platforms. Platforms are not required to assess lawfulness of content. Moderation is typically guided by a platform's terms and conditions to avoid the cost and complexity of legal review \cite{Keller2023Rushing}. As a result, content classified as illegal may neither reflect actual unlawfulness nor capture all illegal content \cite{groesch2025big}. These issues may have affected the illegal content probability inferred from the DSA-TDB. 

Future research could extend our model to compare the effects of takedown versus alternative types of interventions, as well as explore combinations of different intervention instruments. 
Finally, future models should consider the trade-offs between the harm caused by persistent illegal content and the unitended effects of moderation tools, e.g., the limitation of free speech caused by false-positive errors.

\section*{Author contributions}

BT, SK, FM, FS, NJ, SG, and LL conceptualized the study. 
FM, NJ, SG, and LL acquired funding. 
BT, SK, and FM developed the methodology. 
FS and NJ analyzed the regulatory context and data sources.
BT, SK, ES, EV, and GN performed formal analysis. 
BT and FM developed the software.  
FM provided computing resources. 
BT conducted the investigation. 
BT and SK visualized the data. 
BT, SK, FM, FS and NJ drafted the manuscript. 
All authors reviewed and edited the draft. 

\section*{Data availability}
Code and data to implement the model and reproduce results are available at \url{github.com/osome-iu/simsom_removal}. 

\section*{Acknowledgements}

We are grateful to Alessandro Flammini, Taekho You, Samuel Groesch, and Azza Bouleimen for helpful discussions; and to Nick Liu for the data collection used to estimate user activities. This work was supported in part by the Swiss National Science Foundation (Sinergia grant CRSII5\_209250) and by the Knight Foundation.

\bibliographystyle{unsrt}
\bibliography{main.bib}

@article{
    Kozyreva2023dilemmas,
    author = {Anastasia Kozyreva  and Stefan M. Herzog  and Stephan Lewandowsky  and Ralph Hertwig  and Philipp Lorenz-Spreen  and Mark Leiser  and Jason Reifler },
    title = {Resolving content moderation dilemmas between free speech and harmful misinformation},
    journal = {Proceedings of the National Academy of Sciences},
    volume = {120},
    number = {7},
    pages = {e2210666120},
    year = {2023},
    doi = {10.1073/pnas.2210666120},
    URL = {https://www.pnas.org/doi/abs/10.1073/pnas.2210666120},
    eprint = {https://www.pnas.org/doi/pdf/10.1073/pnas.2210666120},
}

@misc{Nikolov2020dataset,
    author = {Dimitar Nikolov and Alessandro Flammini and Filippo Menczer},
    howpublished = {Harvard Dataverse},
    title = {{Replication Data for: Right and left, partisanship predicts vulnerability to misinformation}},
    Note = {doi:10.7910/DVN/6CZHH5},
    UNF = {UNF:6:Avb9rregQyEFA8q65GDGWA==},
    year = {2020},
    version = {V2},
    doi = {10.7910/DVN/6CZHH5},
    url = {https://doi.org/10.7910/DVN/6CZHH5}
}

@article{vazquez03Growing,
	Author = {V\'azquez, Alexei},
	Doi = {10.1103/PhysRevE.67.056104},
	Issue = {5},
	Journal = {Phys. Rev. E},
	Numpages = {15},
	Pages = {056104},
	Title = {Growing network with local rules: Preferential attachment, clustering hierarchy, and degree correlations},
	Url = {https://link.aps.org/doi/10.1103/PhysRevE.67.056104},
	Volume = {67},
	Year = {2003}}

@article{bak2022combining,
  title={Combining interventions to reduce the spread of viral misinformation},
  author={Bak-Coleman, Joseph B and Kennedy, Ian and Wack, Morgan and Beers, Andrew and Schafer, Joseph S and Spiro, Emma S and Starbird, Kate and West, Jevin D},
  journal={Nature Human Behaviour},
  volume={6},
  number={10},
  pages={1372--1380},
  year={2022},
  publisher={Nature Publishing Group UK London}
}

@article{truong2023quantifying,
    author = {Truong, Bao Tran and Lou, Xiaodan and Flammini, Alessandro and Menczer, Filippo},
    title = {Quantifying the vulnerabilities of the online public square to adversarial manipulation tactics},
    journal = {PNAS Nexus},
    volume = {3},
    number = {7},
    pages = {pgae258},
    year = {2024},
    month = {06},
    issn = {2752-6542},
    doi = {10.1093/pnasnexus/pgae258},
    url = {https://doi.org/10.1093/pnasnexus/pgae258},
    eprint = {https://academic.oup.com/pnasnexus/article-pdf/3/7/pgae258/58645601/pgae258.pdf},
}

@misc{twitteralgo, 
    title={Twitter's Recommendation Algorithm},  
    howpublished={\url{blog.twitter.com/engineering/en_us/topics/open-source/2023/twitter-recommendation-algorithm}}, 
    author={Twitter}, 
    year={2023}
}

@misc{dsadatabase, 
    title={{Digital Service Act (DSA) Transparency Database}},  
    howpublished={\url{https://transparency.dsa.ec.europa.eu/}}, 
    author={European Commission}, 
    year={2024}
}

@article{morrow2022emerging,
  title={The emerging science of content labeling: Contextualizing social media content moderation},
  author={Morrow, Garrett and Swire-Thompson, Briony and Polny, Jessica Montgomery and Kopec, Matthew and Wihbey, John P},
  journal={Journal of the Association for Information Science and Technology},
  volume={73},
  number={10},
  pages={1365--1386},
  year={2022},
  publisher={Wiley Online Library}
}

@article{martel2023misinformation,
  title={Misinformation warning labels are widely effective: A review of warning effects and their moderating features},
  author={Martel, Cameron and Rand, David G},
  journal={Current Opinion in Psychology},
  pages={101710},
  year={2023},
  publisher={Elsevier}
}

@article{jahn2023friction,
  title={Friction Interventions to Curb the Spread of Misinformation on Social Media},
  author={Jahn, Laura and Rendsvig, Rasmus K and Flammini, Alessandro and Menczer, Filippo and Hendricks, Vincent F},
  journal={arXiv preprint arXiv:2307.11498},
  year={2023}
}

@article{goldstein2023understanding,
  title={Understanding the (in) effectiveness of content moderation: A case study of {Facebook} in the context of the {US} capitol riot},
  author={Goldstein, Ian and Edelson, Laura and Nguyen, Minh-Kha and Goga, Oana and McCoy, Damon and Lauinger, Tobias},
  journal={arXiv preprint arXiv:2301.02737},
  year={2023}
}

@article{schneider2023effectiveness,
  title={The effectiveness of moderating harmful online content},
  author={Schneider, Philipp J and Rizoiu, Marian-Andrei},
  journal={Proceedings of the National Academy of Sciences},
  volume={120},
  number={34},
  pages={e2307360120},
  year={2023}
}

@article{pennycook2020fighting,
  title={Fighting {COVID-19} misinformation on social media: Experimental evidence for a scalable accuracy-nudge intervention},
  author={Pennycook, Gordon and McPhetres, Jonathon and Zhang, Yunhao and Lu, Jackson G and Rand, David G},
  journal={Psychological Science},
  volume={31},
  number={7},
  pages={770--780},
  year={2020},
  publisher={Sage Publications Sage CA: Los Angeles, CA}
}

@article{pennycook2021shifting,
  title={Shifting attention to accuracy can reduce misinformation online},
  author={Pennycook, Gordon and Epstein, Ziv and Mosleh, Mohsen and Arechar, Antonio A and Eckles, Dean and Rand, David G},
  journal={Nature},
  volume={592},
  number={7855},
  pages={590--595},
  year={2021},
  publisher={Nature Publishing Group UK London}
}

@article{pennycook2022accuracy,
  title={Accuracy prompts are a replicable and generalizable approach for reducing the spread of misinformation},
  author={Pennycook, Gordon and Rand, David G},
  journal={Nature Communications},
  volume={13},
  number={1},
  pages={2333},
  year={2022},
  publisher={Nature Publishing Group UK London}
}

@article{tomalin2023rethinking,
  title={Rethinking online friction in the information society},
  author={Tomalin, Marcus},
  journal={Journal of Information Technology},
  volume={38},
  number={1},
  pages={2--15},
  year={2023},
  publisher={SAGE Publications Sage UK: London, England}
}

@inproceedings{bhuiyan2018feedreflect,
  title={FeedReflect: A tool for nudging users to assess news credibility on {Twitter}},
  author={Bhuiyan, Md Momen and Zhang, Kexin and Vick, Kelsey and Horning, Michael A and Mitra, Tanushree},
  booktitle={Companion of the 2018 ACM conference on computer supported cooperative work and social computing},
  pages={205--208},
  year={2018}
}

@article{broniatowski2022evaluating,
  title={Evaluating the efficacy of {Facebook's} vaccine misinformation content removal policies},
  author={Broniatowski, David A and Gu, Jiayan and Jamison, Amelia M and Abroms, Lorien C},
  journal={Europe PMC},
  year={2022}
}

@article{jhaver2019does,
  title={Does transparency in moderation really matter? User behavior after content removal explanations on {Reddit}},
  author={Jhaver, Shagun and Bruckman, Amy and Gilbert, Eric},
  journal={Proceedings of the ACM on Human-Computer Interaction},
  volume={3},
  number={CSCW},
  pages={1--27},
  year={2019},
  publisher={ACM New York, NY, USA}
}

@article{srinivasan2019content,
  title={Content removal as a moderation strategy: Compliance and other outcomes in the changemyview community},
  author={Srinivasan, Kumar Bhargav and Danescu-Niculescu-Mizil, Cristian and Lee, Lillian and Tan, Chenhao},
  journal={Proceedings of the ACM on Human-Computer Interaction},
  volume={3},
  number={CSCW},
  pages={1--21},
  year={2019},
  publisher={ACM New York, NY, USA}
}

@misc{futurefreespeech2024,
  title     = {Preventing Torrents of Hate or Stifling Free Expression Online?},
  author    = {Future of Free Speech Project},
  year      = {2024},
  institution = {Future of Free Speech},
  url       = {https://futurefreespeech.org/preventing-torrents-of-hate-or-stifling-free-expression-online/},
  note      = {Accessed: 2024-09-22}
}

@article{yar2018failure,
  title={A failure to regulate? The demands and dilemmas of tackling illegal content and behaviour on social media},
  author={Yar, Majid},
  journal={International Journal of Cybersecurity Intelligence \& Cybercrime},
  volume={1},
  number={1},
  pages={5--20},
  year={2018}
}

@article{jain2020illegal,
  title={Illegal content monitoring on social platforms},
  author={Jain, Tarun and Hazra, Jishnu and Cheng, TC Edwin},
  journal={Production and Operations Management},
  volume={29},
  number={8},
  pages={1837--1857},
  year={2020},
  publisher={SAGE Publications Sage CA: Los Angeles, CA}
}

@article{jiang2023trade,
  title={A trade-off-centered framework of content moderation},
  author={Jiang, Jialun Aaron and Nie, Peipei and Brubaker, Jed R and Fiesler, Casey},
  journal={ACM Transactions on Computer-Human Interaction},
  volume={30},
  number={1},
  pages={1--34},
  year={2023},
  publisher={ACM New York, NY}
}

@article{gillespie2022not,
  title={Do not recommend? Reduction as a form of content moderation},
  author={Gillespie, Tarleton},
  journal={Social Media+ Society},
  volume={8},
  number={3},
  pages={20563051221117552},
  year={2022},
  publisher={SAGE Publications Sage UK: London, England}
}

@article{macdonald2024moderating,
  title={Moderating borderline content while respecting fundamental values},
  author={Macdonald, Stuart and Vaughan, Katy},
  journal={Policy \& Internet},
  volume={16},
  number={2},
  pages={347--361},
  year={2024},
  publisher={Wiley Online Library}
}

@misc{fbdsa24oct, 
    title={{Regulation (EU) 2022/2065} {Digital Services Act Transparency Report for Facebook}},  
    howpublished={\url{https://transparency.meta.com/sr/dsa-transparency-report-sep2024-facebook}}, 
    author={Facebook}, 
    year={2024}
}

@article{wagner2024mapping,
  title={Mapping Interpretations of the Law in Online Content Moderation in {Germany}},
  author={Wagner, Ben and Kettemann, Matthias C and Tiedeke, Anna Sophia and Rachinger, Felicitas and Sekwenz, Marie-Therese},
  journal={Available at SSRN 4673858},
  year={2024}
}

@misc{vie_publique_2020,
  title        = {Loi du 24 juin 2020 visant à lutter contre les contenus haineux sur internet},
  author       = {{Vie publique}},
  year         = {2020},
  howpublished = {Vie publique, \url{https://www.vie-publique.fr/loi/268070-loi-avia-lutte-contre-les-contenus-haineux-sur-internet}},
  note         = {Accessed: 9 January 2025}
}

@misc{heldt_2020,
  author       = {Heldt, Amélie P.},
  title        = {{Loi Avia: Frankreichs Verfassungsrat kippt Gesetz gegen Hass im Netz}},
  howpublished = {JuWissBlog Nr. 96/2020, \url{https://www.juwiss.de/96-2020/}},
  year         = {2020},
  note         = {Accessed: 9 January 2025}
}

@misc{conseil_constitutionnel_2020,
  author       = {{Conseil Constitutionnel}},
  title        = {Décision n° 2020-801 DC du 18 juin 2020: Loi visant à lutter contre les contenus haineux sur internet, \url{https://www.conseil-constitutionnel.fr/decision/2020/2020801DC.htm}},
  year         = {2020},
  note         = {Accessed: 9 January 2025}
}

@article{jahanbakhsh2021exploring,
  title={Exploring lightweight interventions at posting time to reduce the sharing of misinformation on social media},
  author={Jahanbakhsh, Farnaz and Zhang, Amy X and Berinsky, Adam J and Pennycook, Gordon and Rand, David G and Karger, David R},
  journal={Proceedings of the ACM on Human-Computer Interaction},
  volume={5},
  number={CSCW1},
  pages={1--42},
  year={2021},
  publisher={ACM New York, NY, USA}
}

@article{chandrasekharan2017you,
  title={You can't stay here: The efficacy of reddit's 2015 ban examined through hate speech},
  author={Chandrasekharan, Eshwar and Pavalanathan, Umashanthi and Srinivasan, Anirudh and Glynn, Adam and Eisenstein, Jacob and Gilbert, Eric},
  journal={Proceedings of the ACM on Human-Computer Interaction},
  volume={1},
  number={CSCW},
  pages={1--22},
  year={2017},
  publisher={ACM New York, NY, USA}
}

@inproceedings{ali2021understanding,
  title={Understanding the effect of deplatforming on social networks},
  author={Ali, Shiza and Saeed, Mohammad Hammas and Aldreabi, Esraa and Blackburn, Jeremy and De Cristofaro, Emiliano and Zannettou, Savvas and Stringhini, Gianluca},
  booktitle={Proceedings of the 13th ACM Web Science Conference 2021},
  pages={187--195},
  year={2021}
}

@article{jhaver2021evaluating,
  title={Evaluating the effectiveness of deplatforming as a moderation strategy on {Twitter}},
  author={Jhaver, Shagun and Boylston, Christian and Yang, Diyi and Bruckman, Amy},
  journal={Proceedings of the ACM on Human-Computer Interaction},
  volume={5},
  number={CSCW2},
  pages={1--30},
  year={2021},
  publisher={ACM New York, NY, USA}
}

@article{saurwein2025beyond,
  title     = {Beyond Removal: Visibility Reduction in Content Moderation},
  author    = {Saurwein, Florian and Birrer, Alena and Groesch, Samuel and Just, Natascha},
  journal   = {TechREG Chronicle},
  year      = {2025},
  month     = apr,
  pages     = {35--41},
  publisher = {Competition Policy International}
}

@misc{schulz2019probleme,
  author    = {Schulz, Wolfgang and Kettemann, Matthias C. and Heldt, Amélie P.},
  title     = {Problems and Potentials of the {NetzDG} --- a Reader with Five {HBI} Expert Opinions},
  year      = {2019},
  publisher = {Verlag Hans-Bredow-Institut},
  series    = {Working Papers of the Hans Bredow Institute},
  number    = {48},
  month     = {November},
  doi       = {10.21241/ssoar.71727},
  howpublished       = {https://doi.org/10.21241/ssoar.71727},
  url       = {https://doi.org/10.21241/ssoar.71727}
}

@article{edelson2025measurement,
  title={Measurement and Metrics for Content Moderation: The Multi-Dimensional Dynamics of Engagement and Content Removal on {Facebook}},
  author={Edelson, Laura and Kovba, Borys and Yershova, Hanna and Botelho, Austin and McCoy, Damon and Lauinger, Tobias},
  journal={Journal of Online Trust and Safety},
  volume={2},
  number={5},
  year={2025}
}

@article{groesch2025big,
  title={Big data, small answers: How the {DSA Transparency Database} falls short of its regulatory objectives},
  author={Groesch, Samuel and Birrer, Alena and Just, Natascha and Saurwein, Florian},
  journal={Telecommunications Policy},
  volume={50},
  number={1},
  doi={10.1016/j.telpol.2025.103088},
  url={https://doi.org/10.1016/j.telpol.2025.103088},
  pages={103088},
  year={2026}
}

@article{trujillo2025dsa,
  title={The {DSA Transparency Database}: Auditing self-reported moderation actions by social media},
  author={Trujillo, Amaury and Fagni, Tiziano and Cresci, Stefano},
  journal={Proceedings of the ACM on Human-Computer Interaction},
  volume={9},
  number={2},
  pages={1--28},
  year={2025},
  publisher={ACM New York, NY, USA}
}

@misc{CCIAEurope2017,
  author       = {CCIA Europe and others},
  title        = {{Germany's Draft Network Enforcement Law} is a threat to freedom of expression, established {EU} law and the goals of the {Commission’s DSM Strategy} -- the {Commission} must take action},
  year         = {2017},
  howpublished = {Open letter},
  note         = {Open letter, 22.05.2017},
  url          = {https://edri.org/files/201705-letter-germany-network-enforcement-law.pd}
}

@misc{Keller2023Rushing,
  author       = {Keller, Daphne},
  title        = {Rushing to Launch the {EU's} Platform Database Experiment},
  year         = {2023},
  howpublished = {Stanford Center for Internet and Society Blog},
  url          = {https://cyberlaw.stanford.edu/blog/2023/07/rushing-launch-eus-platform-database-experiment/},
  note         = {Accessed: 2026-01-09}
}

\newpage
\section*{Appendix}

\setcounter{figure}{0}
\setcounter{section}{0}
\setcounter{table}{0}
\renewcommand{\thefigure}{S\arabic{figure}}
\renewcommand{\thetable}{S\arabic{table}}
\renewcommand{\thesection}{S\arabic{section}}

\subsection*{Alternative platform compliance data}
\label{sec:transparency_reports}

The \textit{Transparency Reports} that platforms are required to publish under the DSA (Art.~15) are a potential starting point to understand how content takedown is practiced across different platforms. 
They include metrics such as ``the median time needed for taking the action.'' 

However, despite formal compliance, reporting practices differ significantly across platforms, preventing thorough comparative analysis. 
Moreover, median turnaround times in transparency reports only cover the time between notification and removal of content. As a result, they omit the substantial period during which problematic content circulates in the network prior to notification. 
Finally, these reports do not provide moderation delays specific to illegal content, and platforms do not allow external access to the underlying data, preventing independent validation.

To overcome these limitations, we instead rely on the DSA Transparency Database (DSA-TDB), as described in the main text. 

\subsection*{Illegal content probability modeling}
\label{sec:illegal_prob}

As there is limited access to data on illegal content, it is difficult to determine how such individual probability of posting illegal content is distributed across users. We consider three alternative distributions of individual-level illegal posting probabilities that all share the same population mean $p=0.01$. 

First, we consider a nearly homogeneous scenario in which all users have similar propensities to post illegal content. In this case, $p$ is drawn from a normal distribution with small variance, $P(i) \sim \mathcal{N}(\mu=0.01, \sigma=0.001)$. Second, we consider a skewed but unimodal distribution that allows for moderate heterogeneity. We model this using a beta distribution, $P(i) \sim i^{\alpha-1}(1-i)^{\beta-1}$, where the mean is given by $\frac{\alpha}{\alpha+\beta}$. We set $\alpha=10$ and $\beta=990$, yielding the same mean probability $p=0.01$. 

Third, we explicitly model population heterogeneity by distinguishing between \emph{high-risk} and \emph{low-risk} accounts. This reflects the intuition that a small fraction of users may repeatedly post severe illegal content (e.g., CSAM, terrorist propaganda, or scams), while the majority of users only occasionally violate the law, for example by sharing copyrighted material. In this two-group model, a fraction $s_H$ of users belong to the high-risk group, while the remaining $1-s_H$ belong to the low-risk group. Within each group, individual probabilities are again drawn from beta distributions as described above. 
The average illegal content probability across the full population is thus $p = s_{H} \frac{\alpha_H}{\alpha_H + \beta_H} + (1-s_{H}) \frac{\alpha_L}{\alpha_L + \beta_L}$, where the $\alpha$ and $\beta$ subscripts indicate the groups. We set $s_{H}=0.1$ to reflect a minority of high-risk accounts. We use the values of $\alpha_L=0.1$, $\beta_L=90$ for the low-risk group and  $\alpha_H=3$, $\beta_H=30$ for the high-risk group, resulting in $p=0.01$. 

\begin{figure}
\centering
\includegraphics[width=0.6\linewidth]{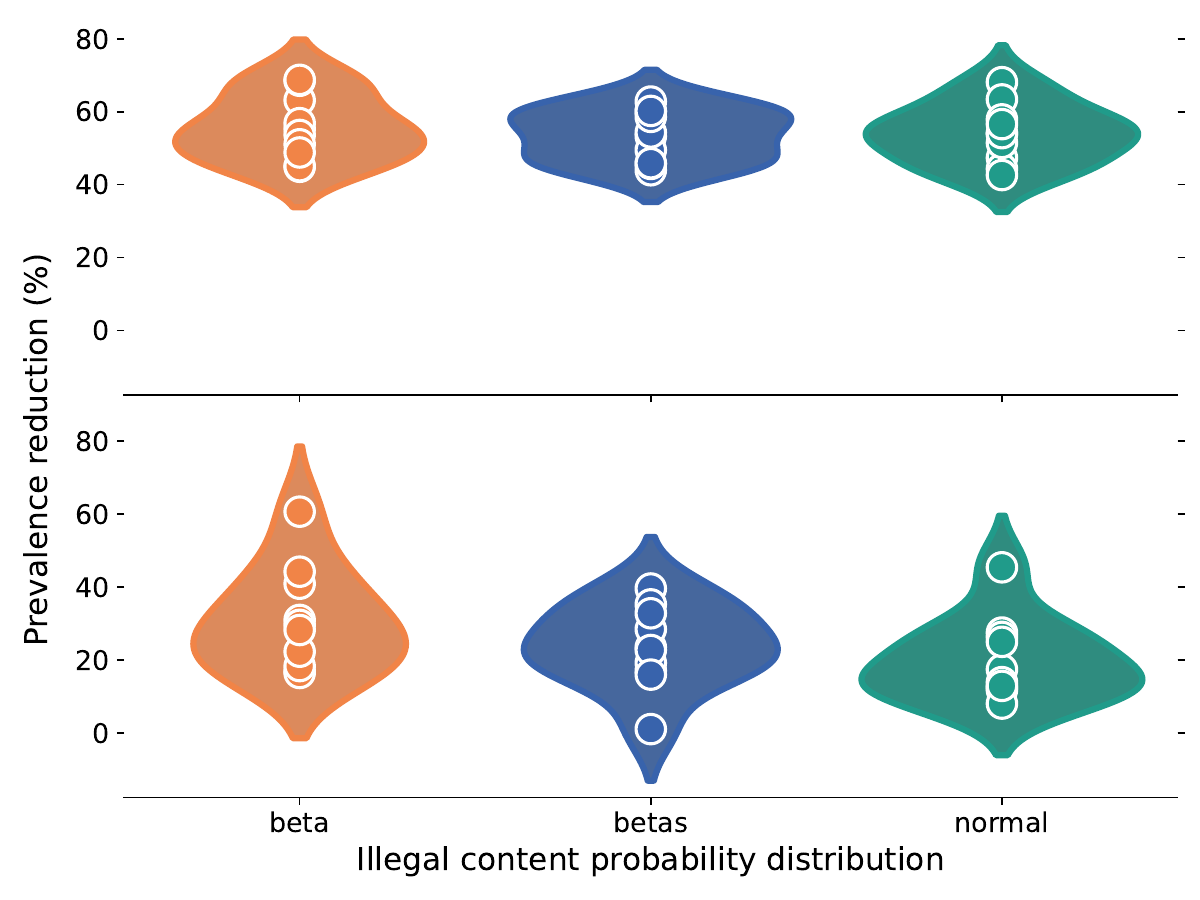}
\caption{Robustness of results with respect to different distributions of $p$ for $\tau=2.89$ (median takedown delay of two days, top) and $\tau=11.54$ (median takedown delay of eight days, bottom) based on 10 simulations. The illegal content reduction does not vary significantly ($P \ge 0.13$ across pairs of simulations with different distributions, using Mann–Whitney U tests with Bonferroni correction).}
\label{fig:distr}
\end{figure}

We explore whether the main outcome variable is sensitive to these different assumptions about illegal content distributions. The reduction in illegal content prevalence shows no statistically significant differences, suggesting that the choice of underlying distribution does not affect the main findings (see Fig.~\ref{fig:distr} in \nameref{sec:robustness}). The simulations discussed in the main text use the two beta distributions. We additionally show that our findings are robust to the relative sizes of the groups (see Fig.~\ref{fig:s_H} in \nameref{sec:robustness}).

\subsection*{Robustness}
\label{sec:robustness}

The results presented in the main text may depend on the distributions of individual-level illegal posting probabilities, the proportion of high-risk accounts $s_H$, the illegal content probability $p$, and the underlying follower network. 
We therefore investigate if the results remain valid when changing these parameters. For these robustness checks, we use two different values of expected content takedown delay, $\tau=2.89$ and $\tau=11.54$, corresponding to median delays of 2 and 8 days, respectively, and assign users to the high- and low-risk groups randomly in each run. 

Fig.~\ref{fig:distr} shows that the main results are robust with respect to different assumptions about the distributions of individual illegal posting probabilities.

For the following analyses varying high-risk group size and illegal content probability, we need a large follower network to generate stable interaction dynamics using low values of these parameters ($s_H=10^{-3}$ and $p=3 \times 10^{-4}$). 
Therefore, we use the original network but remove a random subset of edges to speed up the simulations. The average in/out-degree is thus reduced to 18 friends per node, resulting in a network with $N = 58{,}296$ nodes and $E=1{,}063{,}202$ edges.

\begin{figure}
\centering
\includegraphics[width=0.6\linewidth]{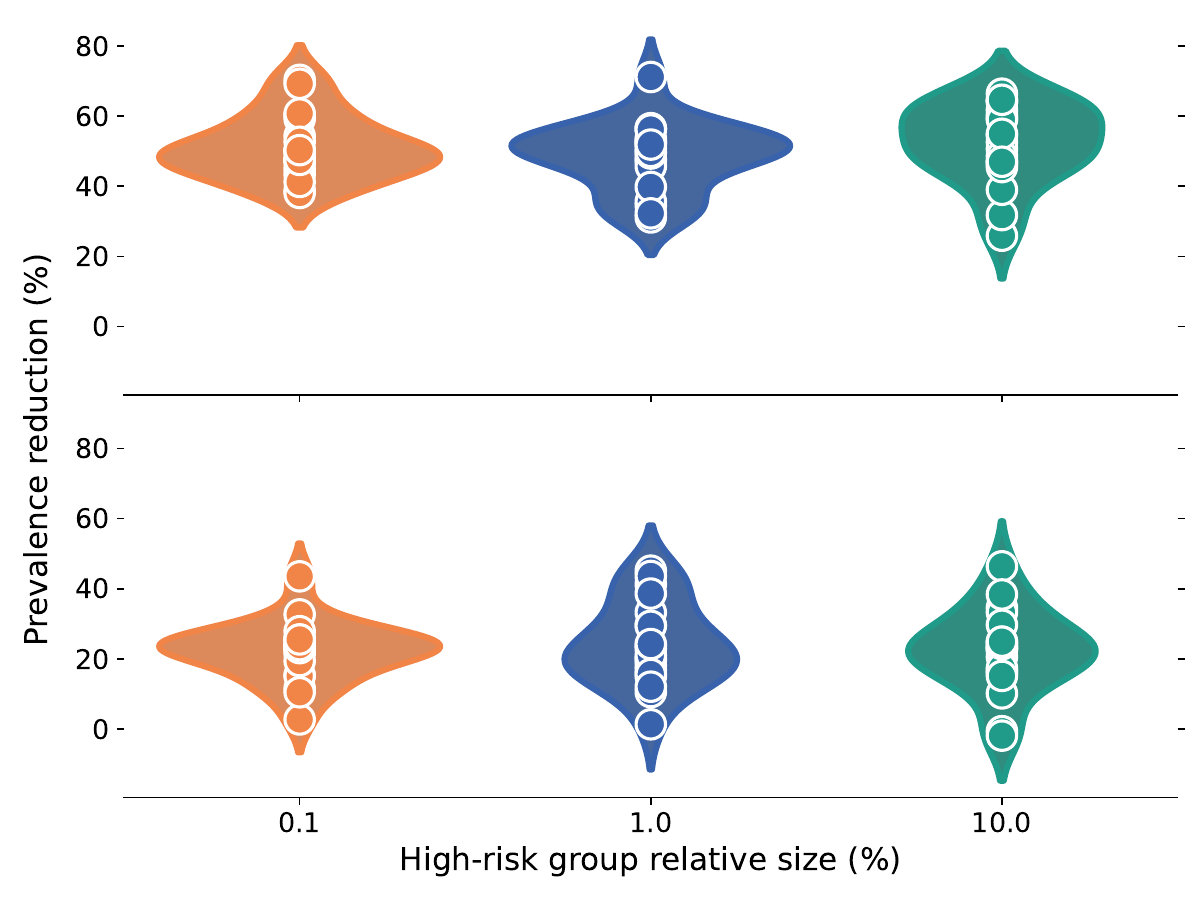}
\caption{Robustness of results with respect to different high-risk group sizes $s_H$  for $\tau=2.89$ (median takedown delay of two days, top) and $\tau=11.54$ (median takedown delay of eight days, bottom) based on 20 simulations. The illegal content reduction does not vary significantly ($P \ge 0.72$ across pairs of simulations with different $s_H$ values, using Mann–Whitney U tests with Bonferroni correction).}
\label{fig:s_H}
\end{figure}

Let us examine the sensitivity of the results to the size of the high-risk group. While our baseline setting assumes $s_H = 0.1$, we explore several orders of magnitude smaller values of the high-risk group size ($10^{-3} \le s_H \le 10^{-1}$) to capture scenarios in which illegal content production is concentrated among a very small fraction of users. 
The main outcome variable, reduction in illegal content prevalence, shows no statistically significant differences (Fig.~\ref{fig:s_H}), suggesting that our findings are robust to changes in $s_H$. 

\begin{figure}
\centering
\includegraphics[width=\linewidth]{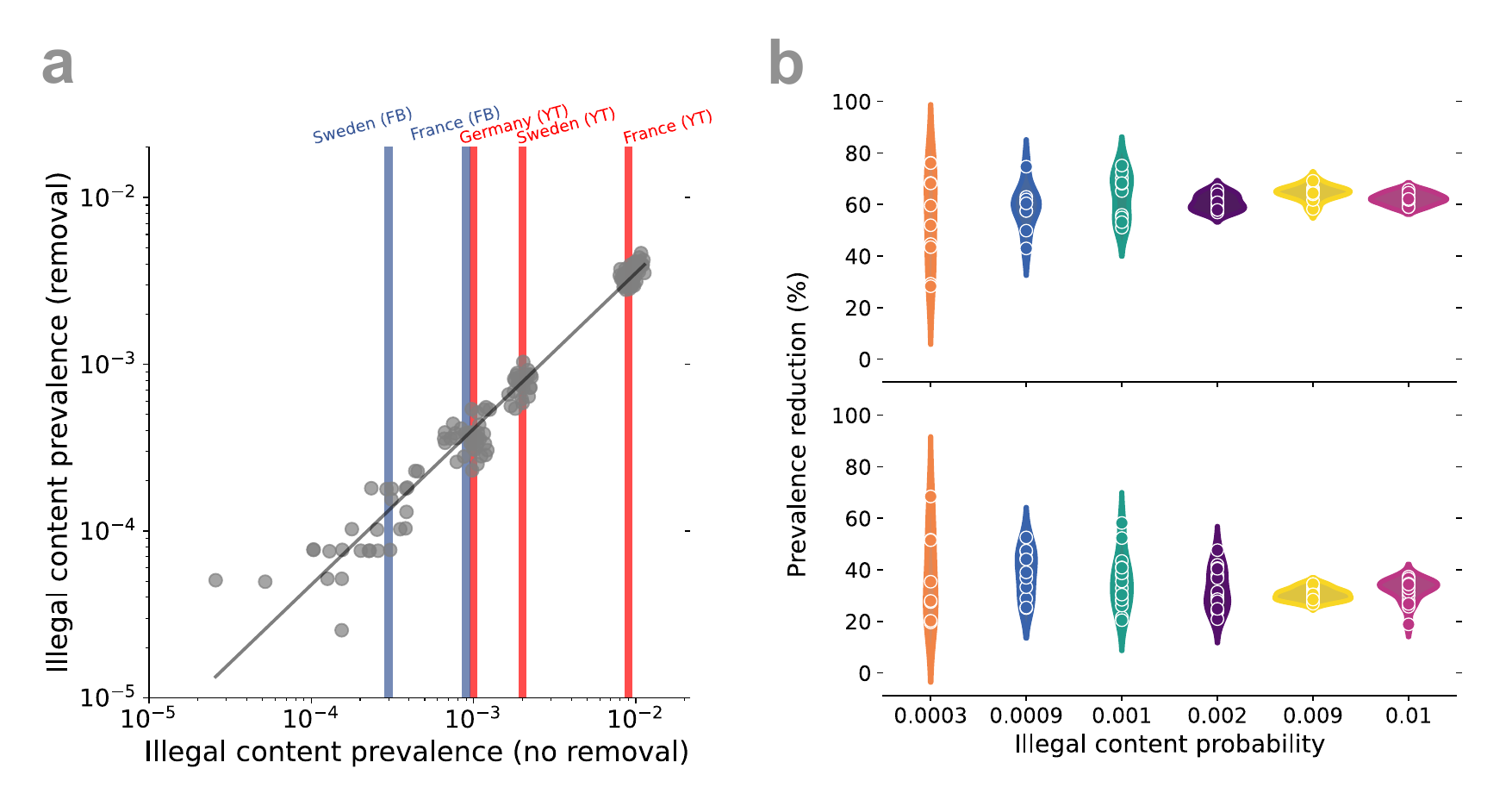}
\caption{Robustness analysis for illegal content probability. 
(a)~Scaling between illegal content prevalence in removal and baseline conditions. Each data point represents the proportion of illegal content in the system with and without removal. 
Results are based on 40 simulations using $\tau=2.89$ (median takedown delay of two days). 
Each simulation is done with an empirical estimate of $p$ (vertical lines) for Facebook and YouTube users from Germany, France, and Sweden, derived from the data reported in Table~\ref{table:futurefreespeech_illegalprob}. 
Estimates of $p$ are based on the proportion of disappeared content that is subsequently judged to be legal. 
The gray line represents a least-squares fit of the log-transformed data, yielding a slope of $0.94 \pm 0.01$. 
(b)~Illegal content reduction for different illegal content probabilities corresponding to the same countries and platforms, based on 10 simulations with $\tau=2.89$ (median takedown delay of two days, top) and $\tau=11.54$ (median takedown delay of eight days, bottom). The result does not vary significantly ($P \ge 0.13$ across pairs of simulations with different $p$ values, using Mann–Whitney U tests with Bonferroni correction).}
\label{fig:illegal_prob}
\end{figure}

We wish to explore how the main result is affected by the volume of illegal content, which varies across platforms and countries. We run simulations with the empirical estimates of $p$ reported in Table~\ref{table:futurefreespeech_illegalprob} (except for the smallest value, which would require a network with infeasibly large size). 
In the main text, we report the reduction in illegal content prevalence, $1 - \frac{\bar{I}_{\text{removal}}}{\langle \bar{I}_{\text{baseline}} \rangle}$. To show that this ratio is independent of $p$, we plot the scaling relationship between $\bar{I}_{\text{removal}}$ and $\bar{I}_{\text{baseline}}$ in Fig.~\ref{fig:illegal_prob}a for different $p$ values from Table~\ref{table:futurefreespeech_illegalprob}. We observe only a weak sublinear relationship, $\bar{I}_{\text{removal}} \sim \bar{I}_{\text{baseline}}^{0.94}$, suggesting that the ratio is nearly independent of $p$. 
We also plot in Fig.~\ref{fig:illegal_prob}b the reduction in the prevalence of illegal content for different values of $p$ from Table~\ref{table:futurefreespeech_illegalprob}. We observe more noise for small $p$, but none of the differences between the values in our main results ($p=0.01$) and those for smaller $p$ values are statistically significant.  
These analyses suggest that our main results are robust with respect to the volume of illegal content in the system. 

\begin{figure}
\centering
\includegraphics[width=0.6\linewidth]{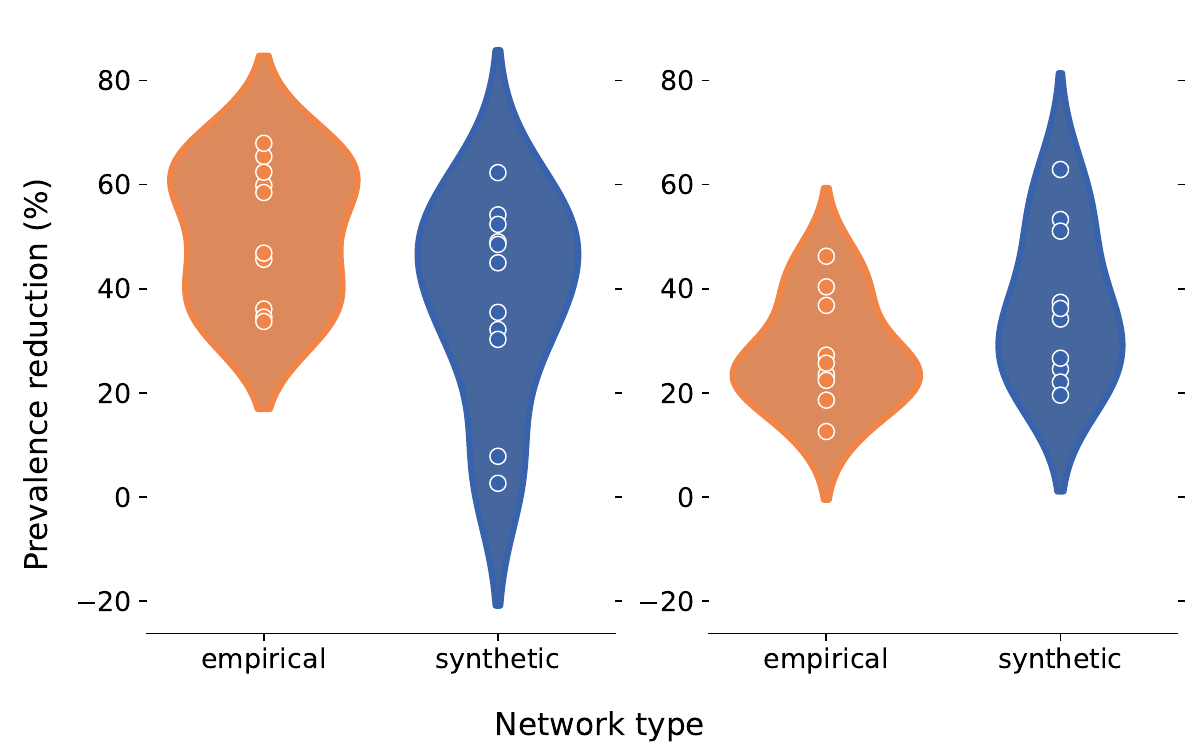}
\caption{Robustness analysis with respect to network structures based on 10 simulations with $\tau=2.89$ (median takedown delay of two days, left) and $\tau=11.54$ (median takedown delay of eight days, right). The illegal content reduction does not vary significantly ($P= 0.11$ and $P=0.22$, respectively, using Mann–Whitney U tests with Bonferroni correction).}
\label{fig:syntheticnet}
\end{figure}

Last, we examine the effect of network structure by comparing simulations using the empirical network and a synthetic network. We create a synthetic network with features ubiquitous in real-world networks, namely, the presence of hubs and clustering (directed triads). To this end, we use a directed variant of the random-walk growth model~\cite{vazquez03Growing}. 
We configure the parameters of this model to generate a network comparable in size and average degree to the network used in the main experiment. In particular, we initialize the network with 181 fully connected nodes and assume new nodes have fixed out-degree $k_{out}=180$. New nodes are added by first randomly selecting one node to follow. For the remaining  $k_{out} - 1$ connections, each new node has an equal probability to follow a random friend of the initially chosen node or a randomly selected node in the network. This results in a network with $N = 10{,}006$ nodes and $E=1{,}810{,}905$ edges.
There is no statistically significant difference in the reduction in illegal content prevalence between these two networks, indicating the robustness of our findings (Fig.~\ref{fig:syntheticnet}).

\subsection*{Empirical expected takedown delay by illegal content type}
\label{sec:by_type}

\begin{figure}
\centering
\includegraphics[width=\linewidth]{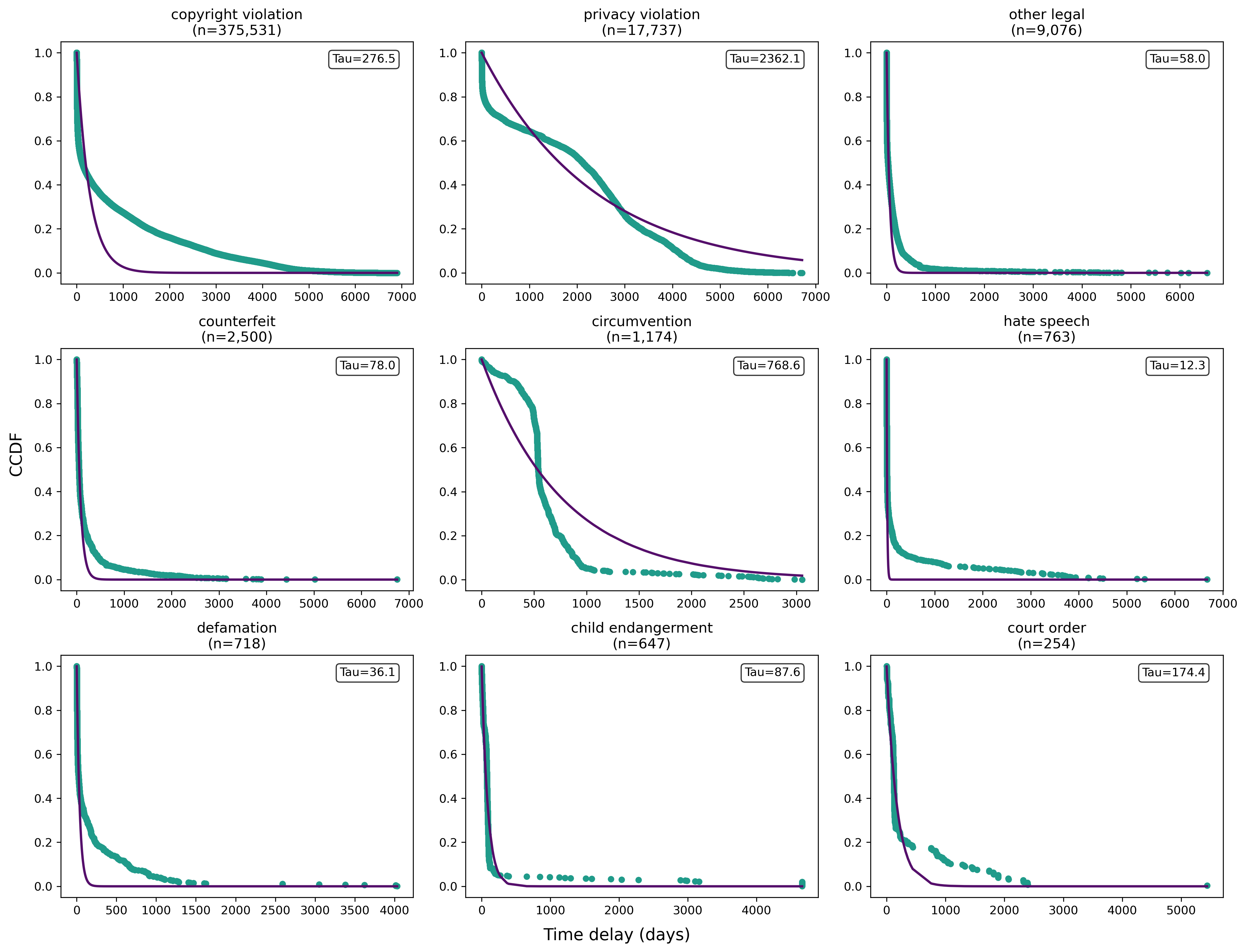}
\caption{Distributions (CCDF) of time delay in takedown on YouTube by type of illegal content. Each plot reports empirical data from the DSA-TDB (green dots) along with an exponential fit (purple line).}
\label{fig:youtube_illegal}
\end{figure} 

\begin{figure}
\centering
\includegraphics[width=\linewidth]{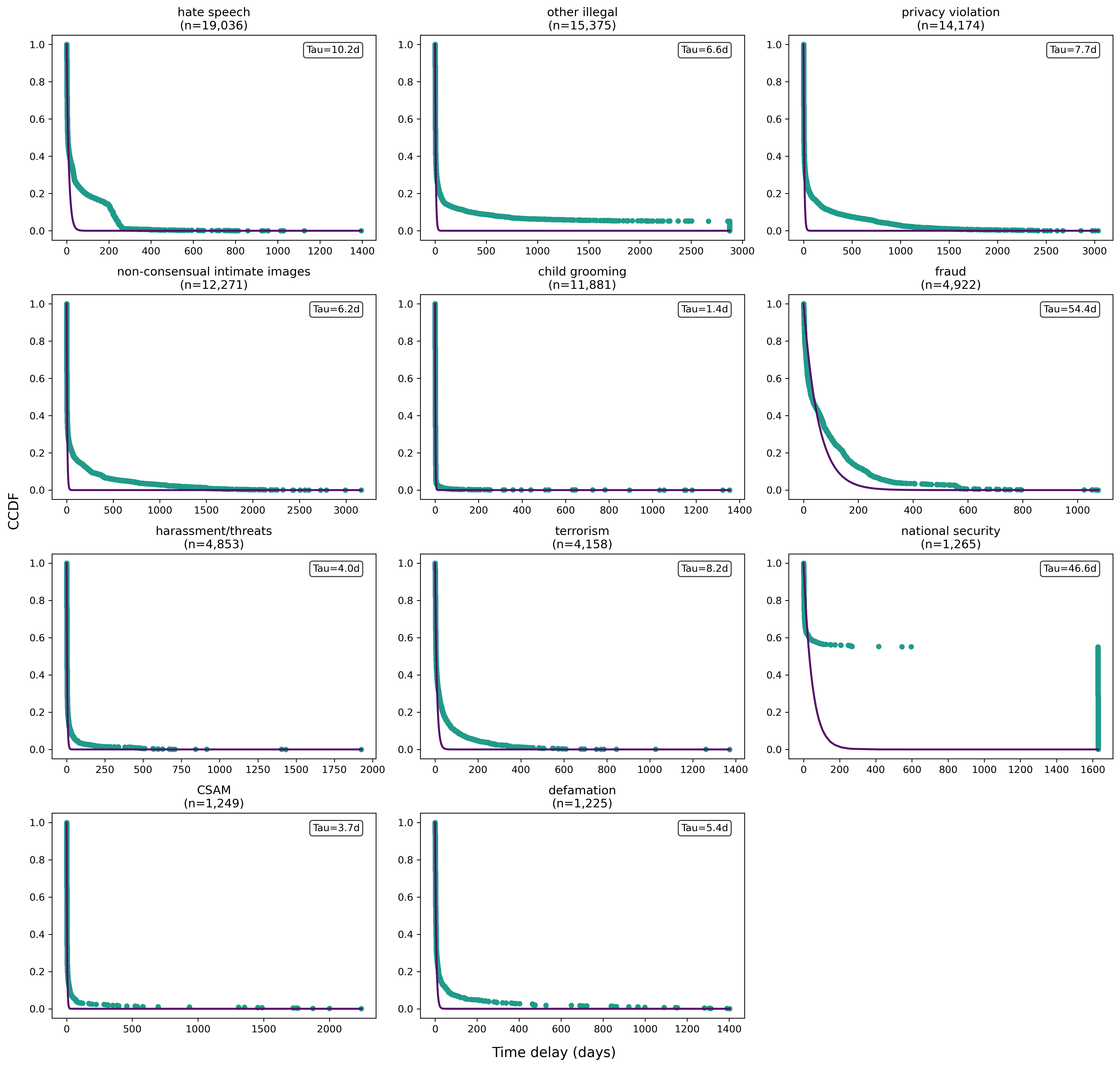}
\caption{Distributions (CCDF) of time delay in takedown on TikTok by type of illegal content. Each plot reports empirical data from the DSA-TDB (green dots) along with an exponential fit (purple line).}
\label{fig:tiktok_illegal}
\end{figure} 

We wish to examine the variation in takedown delays across types of illegal content (Table~\ref{table:illegal_content_delay_platform}). 
Fig.~\ref{fig:youtube_illegal} and Fig.~\ref{fig:tiktok_illegal} show empirical survival time curves for different illegal content types from DSA-TDB SoR records of YouTube and TikTok. By fitting exponential survival time curves $P(t) \sim e^{-t/\tau}$ to this empirical data, we estimate the ranges of $\tau$ for these platforms: 12 days $\leq \tau \leq$ 2,362 days for YouTube and 1 day $\leq  \tau \leq$ 54 days for TikTok.

\end{document}